\documentclass[reprint,showpacs,preprintnumbers,pre,superscriptaddress]{revtex4-1}
\usepackage{graphicx}
\usepackage{amssymb}
\usepackage{amsmath}
\usepackage{amsfonts,color}
\usepackage[T1]{fontenc}
\begin{document}

\title{Stochastic thermodynamics with odd controlling parameters}
\author{Geng Li}
\affiliation{Department of Physics, Beijing Normal University, Beijing 100875, China}
\affiliation{CAS Key Laboratory for Theoretical Physics, Institute of Theoretical Physics, Chinese Academy of Sciences, Beijing 100190, China}
\author{Z. C. Tu}\email[Corresponding author. Email: ]{tuzc@bnu.edu.cn}
\affiliation{Department of Physics, Beijing Normal University, Beijing 100875, China}


\begin{abstract}
Stochastic thermodynamics extends the notions and relations of classical thermodynamics to small systems that experience strong fluctuations. The definitions of work and heat and the microscopically reversible condition are two key concepts in the current framework of stochastic thermodynamics. Herein, we apply stochastic thermodynamics to small systems with odd controlling parameters and find that the definition of heat and the microscopically reversible condition are incompatible. Such a contradiction also leads to a revision to the fluctuation theorems and nonequilibrium work relations. By introducing adjoint dynamics, we find that the total entropy production can be separated into three parts, with two of them satisfying the integral fluctuation theorem. Revising the definitions of work and heat and the microscopically reversible condition allows us to derive two sets of modified nonequilibrium work relations, including the Jarzynski equality, the detailed Crooks work relation, and the integral Crooks work relation. We consider the strategy of shortcuts to isothermality as an example and give a more sophisticated explanation for the Jarzynski-like equality derived from shortcuts to isothermality.
\end{abstract}
\maketitle

\section{Introduction}
In the past several decades, growing interest in small systems has been boosted by the tremendous progress in nanotechnology~\cite{Perkin1994,Liphardt2002,Collin2005,Ciliberto2017} and biomolecular machines~\cite{Balzani2000,Howard2001,Yildiz2004,Kay2007,Millic2014}. Because small systems are susceptible to thermal fluctuations, the mean values of thermodynamic quantities such as work, heat, and entropy are not sufficient to predict the behavior of small systems. Fluctuations and probability distributions of thermodynamic quantities also play a vital role. The framework of stochastic thermodynamics has been developed to study fluctuating behaviors of thermodynamic quantities on individual trajectories~\cite{Sekimoto2010,Jarzynski2011,Seifert2012,Klages2013,Qian2014}. By applying the first law of thermodynamics to fluctuating trajectories, Sekimoto~\cite{Sekimoto1997,Sekimoto1998} first defined work and heat on individual trajectories. A steady-state thermodynamic framework was put forward later by Oono and Paniconi~\cite{Oono1998} and further refined by Hatano and Sasa~\cite{Hatano2001}. However, the generality of the current framework of stochastic thermodynamics remains a fertile topic to be applied to more complex conditions~\cite{Jarzynski2017PRX,Mandal2017}.

The probability distributions of these thermodynamic quantities obey various exact fluctuation relations. These include the Jarzynski equality~\cite{Jarzynski1997,Jarzynski1997PRE,Hummer2001}, which connects free energy differences between two equilibrium states with an exponential average over nonequilibrium work along fluctuating trajectories; the Crooks relation~\cite{Crooks1999PRE,Crooks2000PRE}, which relates the probability distribution of nonequilibrium work in the forward driving processes to the probability distribution of nonequilibrium work in the time-reversed driving processes by means of a detailed and an integral equality; and a series of fluctuation theorems~\cite{Evans1993,GallavottiPRL1995,Evans2002,Seifert2005,Liu2009,Gong2015}, which use the integral and detailed equalities, respectively, to describe the exponential average and the probability distributions of entropy production along fluctuating trajectories. Most of these fluctuation relations can be derived from a fundamental relation: the microscopically reversible condition connecting the probability functionals of the forward and the time-reversed trajectories with the stochastic heat along the forward trajectory~\cite{Crooks1998,Crooks1999,Jarzynski2000J}.

The fluctuation relations listed above mainly focused on the probability distributions of entire thermodynamic quantities. According to the framework of steady-state thermodynamics~\cite{Oono1998,Hatano2001}, the total heat of a stochastic trajectory can be separated into a housekeeping and an excess part. The former is necessary to maintain the system in the nonequilibrium steady state, while the latter is associated with transitions between nonequilibrium steady states. Hatano and Sasa~\cite{Hatano2001} found that the part of the total entropy production related to excess heat satisfies the integral fluctuation theorem. Speck and Seifert~\cite{Speck2005} then demonstrated that the remaining part of the entropy production related to housekeeping heat also satisfies the integral fluctuation theorem. By introducing an adjoint dynamics~\cite{Maes1999,Chernyak2006}, Esposito and Van den Broeck~\cite{Esposito2010PRL,Esposito2010PRE,Broeck2010PRE} generalized the Hatano-Sasa~\cite{Hatano2001} and the Speck-Seifert fluctuation theorem~\cite{Speck2005} into discrete stochastic systems and gave a detailed version of them. Spinney and Ford~\cite{Spinney2012PRL,Spinney2012PRE,Ford2012PRE} recently investigated systems with odd dynamical variables (such as momentum that changes its sign under time-reversal operation) and found that the total entropy production can be separated into three parts, with two of them satisfying the integral fluctuation theorem. Lee et al.~\cite{Lee2013,Yeo2016} modified the separation rule for the total entropy production put forward in~\cite{Spinney2012PRL,Spinney2012PRE,Ford2012PRE} and endowed each part of the total entropy production with clear physical origins.

The controlling parameters, which control small systems during stochastic processes, are usually assumed to be even variables. However, in recent studies we found that controlling parameters in small systems can also be odd variables. One example is from controlling theories in small systems. Shortcut to isothermality provides a unified controlling strategy for conducting a finite-rate isothermal transition between equilibrium states with the same temperature~\cite{Geng2017}. An auxiliary potential is introduced in this strategy to escort the evolution of small systems. It has been demonstrated that the auxiliary potential in shortcuts to isothermality always contains the time derivative of an even controlling parameter. As with the velocity variable, which is the time derivative of the position variable, the time derivative of the even controlling parameter should also change its sign under time-reversal operation, thus failing to guarantee the strategy of shortcuts to isothermality in the time-reversed driving process. In other controlling theories~\cite{Adib2005,Vaikuntanathan2008,Ballard2009}, the time derivative of the even controlling parameter is also contained in the external field, which results in the failure of the controlling theory in the time-reversed driving process. Active biological systems can be considered as another example. Mandal et al.~\cite{Mandal2017} studied entropy production and fluctuation theorems for active matter. The effective field in their formalism also contains the time derivative of an even controlling parameter.  Moreover, the applied magnetic field in charged Brownian particle systems also change its sign under time-reversal operation. To distinguish between different types of controlling parameters, we call the parameters that retain their sign under time-reversal operation as even controlling parameters and those that change their sign under time-reversal operation as odd controlling parameters.

\begin{figure*}[htp]
\centering
 \includegraphics[width = 16cm]{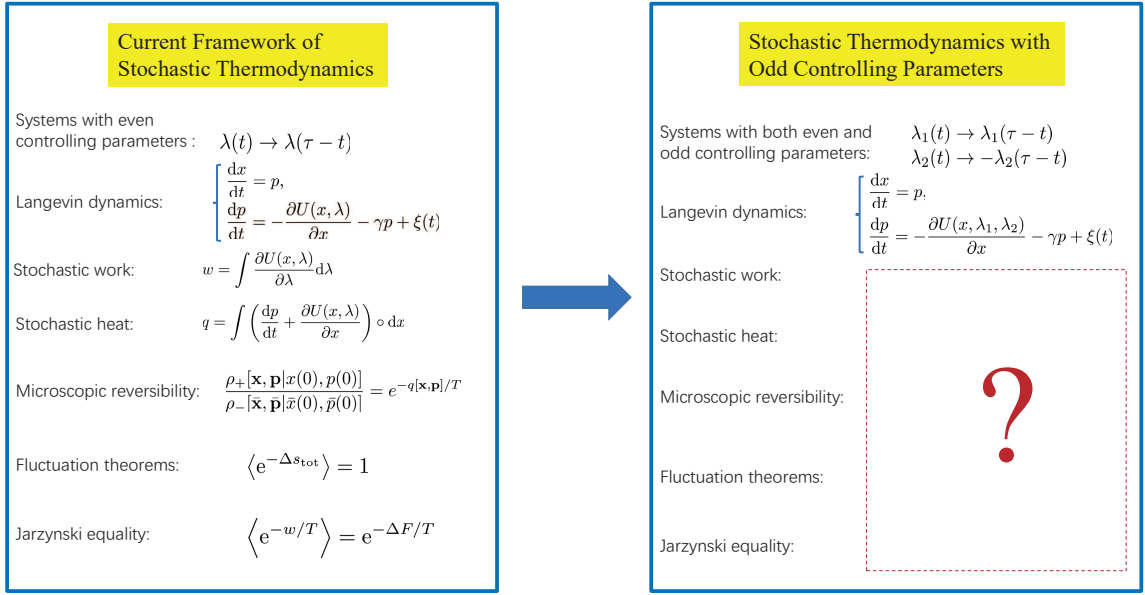}
 \caption{\label{fig1} Schematic of the research motivation of this paper. For the explanation of the meaning of each physical quantity, see the main text.}
\end{figure*}

In this paper, we focus on small systems with both odd and even controlling parameters. However, to highlight the unique properties of odd controlling parameters, we still set the topic as stochastic thermodynamics with odd controlling parameters. As shown in figure~\ref{fig1}, the research motivation of this paper is to determine whether the current framework of stochastic thermodynamics can be extended to systems with odd controlling parameters. If not, can we construct a matched framework of stochastic thermodynamics for systems with odd controlling parameters? In Section~\ref{defwh}, we apply the current framework of stochastic thermodynamics to systems with odd controlling parameters and show that there is a contradiction between the definition of heat and the microscopically reversible condition. Such a contradiction also leads to a revision to the fluctuation theorems and the nonequilibrium work relations. In Section~\ref{enproft}, we discuss the entropy production and fluctuation theorems for systems with odd controlling parameters. By introducing an adjoint dynamics, we can separate the total entropy production into three parts. The total entropy production and two of the three parts satisfy the integral fluctuation theorem. In Section~\ref{nonwore}, we discuss nonequilibrium work relations for systems with odd controlling parameters. We find that two sets of nonequilibrium work relations can be obtained by revising the definitions of work and heat and the microscopically reversible condition, respectively. In Section~\ref{examplesti}, we consider the strategy of shortcuts to isothermality as an example. Starting from two different choices of the definition of work, we give different explanations of the Jarzynski-like equality derived from shortcuts to isothermality. We conclude with a discussion in Section~\ref{disclu}.

\section{Contradiction Between The Definition of Heat and Microscopic Reversibility\label{defwh}}
The general setting considered in this paper involves a Brownian particle coupled to a thermal reservoir with a constant temperature $T$. External driving is applied to the system by introducing a time-dependent potential $U(x,\lambda_{1},\lambda_{2})$ within the time interval $[0 , \tau]$, with $\lambda_{1} \equiv \lambda_{1}(t)$ and $\lambda_{2} \equiv \lambda_{2}(t)$ representing even and odd controlling parameters, respectively. The motion of the stochastic system is governed by the Langevin equation
\begin{equation} \frac{\mathrm{d}x}{\mathrm{d}t}=p, \quad     \frac{\mathrm{d}p}{\mathrm{d}t} = - \frac{\partial U(x,\lambda_{1},\lambda_{2})}{\partial x} - \gamma p + \xi(t), \label{eq:underdampedLE}\end{equation}
where $\gamma$ represents the coefficient of friction and $\xi(t)$ denotes the standard Gaussian white noise satisfying $\langle \xi(t) \rangle=0$ and $\langle \xi(t) \xi(t^\prime) \rangle=2\gamma T \delta(t-t^\prime)$. The ensemble behavior of the stochastic system is described by the Fokker-Planck equation
\begin{equation}   \frac{\partial \rho}{\partial t} = -\frac{\partial}{\partial x} (p\rho) + \frac{\partial}{\partial p}\left( \gamma p\rho + \frac{\partial U}{\partial x}\rho + \gamma T \frac{\partial \rho}{\partial p} \right), \label{eq:underdampedFP}\end{equation}
with $\rho \equiv \rho(x,p,t)$ representing the distribution function of the stochastic system. Throughout this paper, we set the mass of the system and the Boltzmann factor to be unit for simplicity.

Sekimoto proposed to endow the Langevin dynamics with a thermodynamic interpretation~\cite{Sekimoto2010}. The difference between odd and even controlling parameters was not considered in his original formalism. Stochastic work and heat are respectively defined along an individual stochastic trajectory as
\begin{eqnarray}  \text{\dj} w \equiv   \frac{\partial U(x,\lambda_{1},\lambda_{2})}{\partial \lambda_{1}} \mathrm{d} \lambda_{1}+ \frac{\partial U(x,\lambda_{1},\lambda_{2})}{\partial \lambda_{2}}  \mathrm{d} \lambda_{2} , \label{eq:underdampedworkheat1}\end{eqnarray}
and
\begin{eqnarray}  \text{\dj} q  \equiv \left( \frac{\mathrm{d} p}{\mathrm{d}t} + \frac{\partial U(x,\lambda_{1},\lambda_{2})}{\partial x} \right) \circ \mathrm{d} x . \label{eq:underdampedworkheat2}\end{eqnarray}
We adopt the Stratonovich convention $\circ$ in this paper.

The condition of microscopic reversibility
\begin{equation}  \frac{\rho_{+}[\mathbf{x},\mathbf{p}| x(0),p(0)]}{\rho_{-}[\bar{\mathbf{x}},\bar{\mathbf{p}}| \bar{x}(0),\bar{p}(0)]} = e^{- q[\mathbf{x},\mathbf{p}]/T}  \label{eq:microreversibility}\end{equation}
provides another way to define heat~\cite{Crooks1998,Crooks1999,Jarzynski2000J}. Here, $\rho_{+}[\mathbf{x},\mathbf{p}| x(0),p(0)]$ denotes the probability of the trajectory $\{\mathbf{x},\mathbf{p}\}$, given that the system started at $\{x(0),p(0)\}$ and is driven by the protocol $\{ \lambda_{1}(t),\lambda_{2}(t) \}$. The probability of the time-reversed trajectory $\{\bar{\mathbf{x}},\bar{\mathbf{p}}\}$ with the time-reversed dynamics is expressed as $\rho_{-}[\bar{\mathbf{x}},\bar{\mathbf{p}}| \bar{x}(0),\bar{p}(0)]$, given that the system started at $\{\bar{x}(0),\bar{p}(0)\}$ and is driven by the time-reversed protocol $\{ \bar{\lambda}_{1}(\bar{t}),\bar{\lambda}_{2}(\bar{t})\}$. Here $``-"$ represents the time-reversal operation. $q[\mathbf{x},\mathbf{p}]$ denotes the heat absorbed from the thermal reservoir by the system along the forward trajectory.

The mapping relation between the time-reversed trajectory and the forward trajectory reads
\begin{eqnarray}  && \bar{t} = \tau - t,~\bar{x}(\bar{t}) =  x(\tau - \bar{t}),~\bar{p}(\bar{t}) =  -p(\tau - \bar{t}),\nonumber \\ &&\bar{\lambda}_{1}(\bar{t}) = \lambda_{1}(\tau - \bar{t}),~  \bar{\lambda}_{2}(\bar{t}) = -\lambda_{2}(\tau - \bar{t}). \label{eq:undertimereversedvariables}\end{eqnarray}
The time-reversed driving process follows the same dynamical equation as the forward driving process, namely
\begin{equation} \frac{\mathrm{d} \bar{x}}{\mathrm{d}\bar{t}}=\bar{p}, \quad     \frac{\mathrm{d}\bar{p}}{\mathrm{d}\bar{t}} = - \frac{\partial U(\bar{x},\bar{\lambda}_{1},\bar{\lambda}_{2})}{\partial \bar{x}} - \gamma \bar{p} + \xi(\bar{t}), \label{eq:underdampedreversedLE}\end{equation}
while the dynamical variables and the parameters are replaced by the time-reversed ones.

The probability of observing a trajectory of the system for given initial state can be obtained by using the path-integral methods~\cite{Onsager1953,Wiegel1986,Klages2013}. For the forward driving process, the conditional probability of observing increments $\mathrm{d} x = x(t+\mathrm{d}t) - x(t)$ and $\mathrm{d} p = p(t+\mathrm{d}t) - p(t)$ in a time interval $\mathrm{d} t$ can be written as
\begin{eqnarray} && \rho_{+}[x(t+\mathrm{d}t),p(t+\mathrm{d}t)| x(t),p(t)] \nonumber \\&&= \frac{\delta (\mathrm{d} x -p\mathrm{d}t)}{\sqrt{4 \pi \gamma T \mathrm{d}t }} \mathrm{exp} \left [ - \frac{\left ( \mathrm{d}p + \gamma p \mathrm{d}t+   \frac{\partial U(x,\lambda_{1},\lambda_{2})}{\partial x} \mathrm{d}t \right)^{2}}{4 \gamma T \mathrm{d}t}    \right ] . \label{eq:underdampedfstp}\end{eqnarray}
The detailed derivations of this short-time conditional probability are provided in Appendix~\ref{sec:shorttime}. Similarly, the short-time conditional probability of the time-reversed driving process can be expressed as
\begin{eqnarray} &&\rho_{-}[\bar{x}(\bar{t}+\mathrm{d}\bar{t}),\bar{p}(\bar{t}+\mathrm{d}\bar{t})| \bar{x}(\bar{t}),\bar{p}(\bar{t})] \nonumber \\&&= \frac{\delta (\mathrm{d} \bar{x} -\bar{p}\mathrm{d}\bar{t})}{\sqrt{4 \pi \gamma T \mathrm{d}\bar{t} }} \mathrm{exp} \left [ - \frac{\left ( \mathrm{d}\bar{p} + \gamma \bar{p} \mathrm{d}\bar{t}+   \frac{\partial U(\bar{x},\bar{\lambda}_{1},\bar{\lambda}_{2})}{\partial \bar{x}} \mathrm{d}\bar{t} \right)^{2}}{4 \gamma T \mathrm{d}\bar{t}}    \right ] . \label{eq:underdampedrstp}\end{eqnarray}
From microscopic reversibility~(\ref{eq:microreversibility}), we can then give a definition of the heat along a stochastic trajectory as
\begin{widetext}
\begin{eqnarray} \text{\dj} q' \equiv && - T \ln \frac{\rho_{+}[x(t+\mathrm{d}t),p(t+\mathrm{d}t)| x(t),p(t)]}{\rho_{-}[\bar{x}(\bar{t}+\mathrm{d}\bar{t}),\bar{p}(\bar{t}+\mathrm{d}\bar{t})| \bar{x}(\bar{t}),\bar{p}(\bar{t})]} \nonumber \\=&& - T \ln \frac{\rho_{+}[x(t+\mathrm{d}t),p(t+\mathrm{d}t)| x(t),p(t)]}{\rho_{-}[x(t-\mathrm{d}t),-p(t-\mathrm{d}t)| x(t),-p(t)]} \nonumber \\ =&& \frac{1}{4\gamma \mathrm{d}t} \left [ \left ( \mathrm{d}p + \gamma p \mathrm{d}t+   \frac{\partial U(x,\lambda_{1},\lambda_{2})}{\partial x} \mathrm{d}t \right)^{2} - \left ( \mathrm{d}p - \gamma p \mathrm{d}t+   \frac{\partial U(x,\lambda_{1},-\lambda_{2})}{\partial x} \mathrm{d}t \right)^{2}    \right] \nonumber \\ = && \left[  \frac{\mathrm{d} p}{\mathrm{d}t} + \frac{1}{2} \left(  \frac{\partial U}{\partial x} + \frac{\partial U^{\mathrm{r}}}{\partial x}    \right)     \right ] \circ \mathrm{d} x  + \frac{1}{2\gamma} \left[  \frac{\mathrm{d} p}{\mathrm{d}t} + \frac{1}{2} \left(  \frac{\partial U}{\partial x} + \frac{\partial U^{\mathrm{r}}}{\partial x}    \right)  \right]   \left(  \frac{\partial U}{\partial x} - \frac{\partial U^{\mathrm{r}}}{\partial x}    \right)\mathrm{d}t , \label{eq:undermrratioheat}\end{eqnarray}
where we have defined $U^{\mathrm{r}}(x,\lambda_{1},\lambda_{2}) \equiv U(x,\lambda_{1},-\lambda_{2})$. Applying the first law of thermodynamics to a stochastic trajectory, we can then define the stochastic work as
\begin{eqnarray}   \text{\dj} w' \equiv && \mathrm{d}e- \text{\dj}q'  \nonumber \\ = && \frac{\partial U}{\partial \lambda_{1}} \mathrm{d} \lambda_{1}+ \frac{\partial U}{\partial \lambda_{2}}  \mathrm{d} \lambda_{2} +\frac{1}{2} \left(  \frac{\partial U}{\partial x} - \frac{\partial U^{\mathrm{r}}}{\partial x}    \right)     \circ \mathrm{d} x   - \frac{1}{2\gamma} \left[  \frac{\mathrm{d} p}{\mathrm{d}t} + \frac{1}{2} \left(  \frac{\partial U}{\partial x} + \frac{\partial U^{\mathrm{r}}}{\partial x}    \right)  \right]   \left(  \frac{\partial U}{\partial x} - \frac{\partial U^{\mathrm{r}}}{\partial x}    \right)\mathrm{d}t,  \label{eq:undermrratiowork}\end{eqnarray}
\end{widetext}
with $e \equiv p^{2}/ 2+ U$ representing the energy of the system. To distinguish them from previous definitions, we use $q'$ and $w'$ to denote the stochastic heat and work derived from microscopic reversibility~(\ref{eq:microreversibility}), respectively. In the particular situation $U^{\mathrm{r}}(x,\lambda_{1},\lambda_{2}) = U(x,\lambda_{1},\lambda_{2})$, we can easily verify that the definitions of work~(\ref{eq:undermrratiowork}) and heat~(\ref{eq:undermrratioheat}) reduce to Eqs.~(\ref{eq:underdampedworkheat1}) and~(\ref{eq:underdampedworkheat2}), respectively.

Unlike systems containing only even controlling parameters, in which the two definitions of work and heat given above should be consistent~\cite{Sekimoto2010,Crooks1999,Crooks1998}, we find that with the inclusion of odd controlling parameters, the definition given by microscopic reversibility diverges substantially from the one given by the current framework of stochastic thermodynamics. This indicates that either the definitions of work and heat or microscopic reversibility in the current framework of stochastic thermodynamics needs to be modified. Because of the fundamental status in stochastic thermodynamics of the work and heat definitions and of microscopic reversibility, such a contradiction may also lead to a revision to the fluctuation theorems and the nonequilibrium work relations.


\section{Entropy Production and Fluctuation Theorems\label{enproft}}
Having introduced the contradiction in the definitions of work and heat, we next investigate the form of the stochastic entropy production and fluctuation theorems for systems with odd controlling parameters.
\subsection{Total Entropy Production and Fluctuation Theorems}
For a stochastic system evolving in the time interval $[0,\tau]$, the probability ratio of the forward trajectory $\{ \mathbf{x},\mathbf{p} \}$ and time-reversed trajectory $\{ \bar{\mathbf{x}},\bar{\mathbf{p}} \}$ defines the total entropy production
\begin{widetext}
\begin{equation}   \Delta s_{\mathrm{tot}}  \equiv  \ln \frac{\rho_{+}[\mathbf{x},\mathbf{p}]}{\rho_{-}[\bar{\mathbf{x}},\bar{\mathbf{p}}]}   =   \ln \left(  \frac{\rho_{+}(x(0),p(0),0)}{\rho_{-}(\bar{x}(0),\bar{p}(0),0)} \right) +  \ln \frac{\rho_{+}[\mathbf{x},\mathbf{p}| x(0),p(0)]}{\rho_{-}[\bar{\mathbf{x}},\bar{\mathbf{p}}| \bar{x}(0),\bar{p}(0)]},              \label{eq:undertotalentflu}\end{equation}
\end{widetext}
where $\rho_{+}(x(0),p(0),0)$ and $\rho_{-}(\bar{x}(0),\bar{p}(0),0)$ represent the initial distributions of the forward and the time-reversed driving processes, respectively.

It has already been proved that a random variable such as $\Delta s_{\mathrm{tot}}$ obeys an integral fluctuation theorem
\begin{equation} \left \langle   \mathrm{e}^{-\Delta s_{\mathrm{tot}}} \right \rangle= \iint \mathrm{d} \mathbf{x}\mathrm{d} \mathbf{p} \rho_{+}[\mathbf{x},\mathbf{p}] \mathrm{e}^{-\Delta s_{\mathrm{tot}}}=1  \label{eq:undertotalIFT}\end{equation}
if this random variable can be expressed as the ratio of the forward trajectory probability functional $\rho_{+}[\mathbf{x},\mathbf{p}]$ and another transformed trajectory probability functional, such as $\rho_{-}[\bar{\mathbf{x}},\bar{\mathbf{p}}]$. The transformed trajectory and the forward trajectory must have the Jacobian of unity, such as $|\partial \bar{\mathbf{x}}/ \partial \mathbf{x}|=1 $ and $|\partial \bar{\mathbf{p}}/ \partial \mathbf{p}|=1 $ ~\cite{Seifert2005,Esposito2010PRL,Esposito2010PRE,Broeck2010PRE,Yang2019}.

The initial distribution of the time-reversed driving process is chosen as the final distribution of the forward driving process, i.e.,
\begin{equation}  \rho_{-}(\bar{x}(0),\bar{p}(0),0)  = \rho_{+}(x(\tau),p(\tau),\tau). \label{eq:reverseinitial}\end{equation}
Substituting Eqs.~(\ref{eq:underdampedfstp}),~(\ref{eq:underdampedrstp}), and~(\ref{eq:reverseinitial}) into Eq.~(\ref{eq:undertotalentflu}), we can express the total entropy production as
\begin{widetext}
\begin{eqnarray}   \Delta s_{\mathrm{tot}} =&& \ln \left(  \frac{\rho_{+}(x(0),p(0),0)}{\rho_{+}(x(\tau),p(\tau),\tau)} \right)  - \frac{1}{T}\int \left[  \frac{\mathrm{d} p}{\mathrm{d}t} + \frac{1}{2} \left(  \frac{\partial U}{\partial x} + \frac{\partial U^{r}}{\partial x}    \right)     \right ] \circ \mathrm{d} x \nonumber \\&& - \frac{1}{2\gamma T} \int^{\tau}_{0}\left[  \frac{\mathrm{d} p}{\mathrm{d}t} + \frac{1}{2} \left(  \frac{\partial U}{\partial x} + \frac{\partial U^{r}}{\partial x}    \right)  \right]   \left(  \frac{\partial U}{\partial x} - \frac{\partial U^{r}}{\partial x}    \right)\mathrm{d}t.              \label{eq:undertotalentropy}\end{eqnarray}
\end{widetext}
Such a form of the total entropy production is much more complicated than the case containing only even controlling parameters~\cite{Seifert2005,Seifert2012,Esposito2010PRL,Esposito2010PRE,Broeck2010PRE}. Therefore the inclusion of odd controlling parameters also changes the form of the total entropy production, but the fluctuation theorem for the total entropy production still holds.

\subsection{Three Parts of the Total Entropy Production}
To further exploit nonequilibrium properties of the entropy, we introduce an adjoint dynamics
\begin{equation} \frac{\mathrm{d}x}{\mathrm{d}t}=-p, \quad     \frac{\mathrm{d}p}{\mathrm{d}t} = \frac{\partial U(x,\lambda_{1},\lambda_{2})}{\partial x} - \gamma p + \xi(t). \label{eq:underdampedadLE}\end{equation}
Note that such a counterintuitive dynamics is mathematically constructed just to separate the total entropy production into meaningful parts~\cite{Esposito2010PRL,Crooks2000PRE,Chernyak2006}. The short-time conditional probability of this adjoint dynamics satisfies the relation
\begin{eqnarray}&& \frac{\rho^{\mathrm{ad}}[x(t),p(t)|x(t+\mathrm{d}t),p(t+\mathrm{d}t)]}{\rho_{+}[x(t+\mathrm{d}t),p(t+\mathrm{d}t)|x(t),p(t)] } \nonumber \\&&=  \frac{\rho_{\mathrm{st}}(x(t),p(t),\lambda_{1},\lambda_{2})}{\rho_{\mathrm{st}}(x(t+\mathrm{d}t),p(t+\mathrm{d}t),\lambda_{1},\lambda_{2})}, \label{eq:adjointdynamics}\end{eqnarray}
where $\rho_{\mathrm{st}}(x,p,\lambda_{1},\lambda_{2})$ represents the stationary distribution of the forward dynamics~\cite{Maes1999}. We can verify that the adjoint dynamics gives the same stationary distribution and the opposite stationary flux as the forward dynamics.

Based on the adjoint dynamics, we construct two adjoint driving processes $\mathrm{A}$ and $\mathrm{B}$.
The mapping relation between the $\mathrm{A}$ adjoint driving process and the forward driving process reads
\begin{eqnarray}  && \hat{t} = \tau - t,~\hat{x}(\hat{t}) =  x(\tau - \hat{t}),~\hat{p}(\hat{t}) =  p(\tau - \hat{t}),\nonumber \\  &&\hat{\lambda}_{1}(\hat{t}) = \lambda_{1}(\tau - \hat{t}),~  \hat{\lambda}_{2}(\hat{t}) = \lambda_{2}(\tau - \hat{t}), \label{eq:underAmaprela}\end{eqnarray}
while the mapping relation between the $\mathrm{B}$ adjoint driving process and the forward driving process reads
\begin{eqnarray}  &&\tilde{t} = t,~\tilde{x}(\tilde{t}) =  x( t),~\tilde{p}(\tilde{t}) = - p( t),\nonumber \\  &&\tilde{\lambda}_{1}(\tilde{t}) = \lambda_{1}( t),~  \tilde{\lambda}_{2}(\tilde{t}) = -\lambda_{2}(t). \label{eq:underBmaprela}\end{eqnarray}
Herein, variables with hat and tilde represent the variables in the $\mathrm{A}$ adjoint driving process and in the $\mathrm{B}$ adjoint driving process, respectively.
The initial distributions of the $\mathrm{A}$ adjoint driving process and the $\mathrm{B}$ adjoint driving process are respectively chosen as
\begin{equation}  \rho_{\mathrm{A}}(\hat{x}(0),\hat{p}(0),0)  = \rho_{+}(x(\tau),p(\tau),\tau) \label{eq:Ainitial}\end{equation}
and
\begin{equation}  \rho_{\mathrm{B}}(\tilde{x}(0),\tilde{p}(0),0)  = \rho_{+}(x(0),p(0),0). \label{eq:Binitial}\end{equation}
By utilizing these two adjoint driving processes, we can further separate the total entropy production into three parts
\begin{equation} \Delta s_{\mathrm{tot}} = \Delta s_{\mathrm{A}} + \Delta s_{\mathrm{B}}+\Delta s_{\mathrm{C}}, \label{eq:separatethree}\end{equation}
with the trajectory entropy related to the $\mathrm{A}$ adjoint driving process
\begin{widetext}
\begin{equation}  \Delta s_{\mathrm{A}} =  \ln \frac{\rho_{+}[\mathbf{x},\mathbf{p}]}{\rho_{A}[\hat{\mathbf{x}},\hat{\mathbf{p}}]} =   \ln \left(  \frac{\rho_{+}(x(0),p(0),0)}{\rho_{\mathrm{A}}(\hat{x}(0),\hat{p}(0),0)} \right)   +   \ln \left( \frac{\rho_{+}[\mathbf{x},\mathbf{p}| x(0),p(0)]}{\rho_{\mathrm{A}}[\hat{\mathbf{x}},\hat{\mathbf{p}}| \hat{x}(0),\hat{p}(0)]} \right), \label{eq:underAdrivingentropy}\end{equation}
the trajectory entropy related to the $\mathrm{B}$ adjoint driving process
\begin{equation} \Delta s_{\mathrm{B}} =  \ln \frac{\rho_{+}[\mathbf{x},\mathbf{p}]}{\rho_{B}[\tilde{\mathbf{x}},\tilde{\mathbf{p}}]} =   \ln \left(  \frac{\rho_{+}(x(0),p(0),0)}{\rho_{\mathrm{B}}(\tilde{x}(0),\tilde{p}(0),0)} \right)   +   \ln \left( \frac{\rho_{+}[\mathbf{x},\mathbf{p}| x(0),p(0)]}{\rho_{\mathrm{B}}[\tilde{\mathbf{x}},\tilde{\mathbf{p}}| \tilde{x}(0),\tilde{p}(0)]} \right) , \label{eq:underBdrivingentropy}\end{equation}
and the remaining part
\begin{equation} \Delta s_{\mathrm{C}} =  \ln \left(\frac{ \rho_{\mathrm{A}}(\hat{x}(0),\hat{p}(0),0) \rho_{\mathrm{B}}(\tilde{x}(0),\tilde{p}(0),0)}{\rho_{+}(x(0),p(0),0)  \rho_{-}(\bar{x}(0),\bar{p}(0),0) }  \right)    +  \ln \left(   \frac{ \rho_{\mathrm{A}}[\hat{\mathbf{x}},\hat{\mathbf{p}}| \hat{x}(0),\hat{p}(0)]  \rho_{\mathrm{B}}[\tilde{\mathbf{x}},\tilde{\mathbf{p}}| \tilde{x}(0),\tilde{p}(0)] }{ \rho_{+}[\mathbf{x},\mathbf{p}| x(0),p(0)]   \rho_{-}[\bar{\mathbf{x}},\bar{\mathbf{p}}| \bar{x}(0),\bar{p}(0)] }  \right).  \label{eq:underCdrivingentropy}\end{equation}
\end{widetext}

According to the structure of the trajectory entropy~(\ref{eq:underAdrivingentropy}) and~(\ref{eq:underBdrivingentropy}), we can prove that $\Delta s_{\mathrm{A}}$ and $\Delta s_{\mathrm{B}}$ satisfy the integral fluctuation theorems
\begin{equation} \left \langle   \mathrm{e}^{-\Delta s_{\mathrm{A}}} \right \rangle=1, ~~~ \left \langle   \mathrm{e}^{-\Delta s_{\mathrm{B}}} \right \rangle=1 . \label{eq:underadjointft}\end{equation}
The third trajectory entropy $\Delta s_{\mathrm{C}}$ does not satisfy the fluctuation theorem since it cannot be written in the form of $\Delta s_{\mathrm{A}}$, $\Delta s_{\mathrm{B}}$, and $\Delta s_{\mathrm{tot}}$.
\begin{widetext}
The concrete expressions of $\Delta s_{\mathrm{A}}$, $\Delta s_{\mathrm{B}}$, and $\Delta s_{\mathrm{C}}$ can be further derived respectively as
\begin{equation} \Delta s_{\mathrm{A}}   = \ln \left (  \frac{\rho_{+}(x(0),p(0),0)}{\rho_{+}(x(\tau),p(\tau),\tau)} \right)  -  \frac{1}{T}\int \left(  \frac{\mathrm{d} p}{\mathrm{d}t} +   \frac{\partial U}{\partial x}      \right ) \circ \mathrm{d} x       ,\label{eq:underAenexpr}\end{equation}
\begin{eqnarray} \Delta s_{\mathrm{B}}   =   -\frac{1}{2T} \int \left( \frac{\partial U}{\partial x}  -  \frac{\partial U^{\mathrm{r}}}{\partial x}     \right)    \circ \mathrm{d} x -  \frac{1}{2\gamma T}   \int^{\tau}_{0} \left [    \frac{\mathrm{d} p}{\mathrm{d}t}   +  \frac{1}{2} \left( \frac{\partial U}{\partial x}  +  \frac{\partial U^{\mathrm{r}}}{\partial x}     \right)     \right ]   \left( \frac{\partial U}{\partial x}  -  \frac{\partial U^{\mathrm{r}}}{\partial x}     \right) \mathrm{d} t ,   \label{eq:underBenexpr}\end{eqnarray}
and
\begin{equation} \Delta s_{\mathrm{C}}   = \frac{1}{T} \int \left(  \frac{\partial U}{\partial x}  -  \frac{\partial U^{\mathrm{r}}}{\partial x}      \right) \circ \mathrm{d} x       .\label{eq:underCenexpr}\end{equation}
\end{widetext}
The detailed derivations of the above expressions are shown in Appendix~\ref{sec:expression}. Note that if the potential satisfies $U^{\mathrm{r}}(x,\lambda_{1},\lambda_{2}) = U(x,\lambda_{1},\lambda_{2})$, we can easily verify that $\Delta s_{\mathrm{B}}=\Delta s_{\mathrm{C}}=0$ and the total entropy production reduces to the case for systems containing only even controlling parameters~\cite{Seifert2005,Seifert2012}. The inclusion of odd controlling parameters leads to two extra entropy contributions $\Delta s_{\mathrm{B}}$ and $\Delta s_{\mathrm{C}}$ and fluctuation theorems~(\ref{eq:underadjointft}).

\section{Nonequilibrium Work Relations\label{nonwore}}
In section~\ref{defwh}, we put forward the contradiction between the definition of heat and microscopic reversibility for stochastic systems with odd controlling parameters. In this section, we revise the definitions of work and heat and microscopic reversibility in the current framework of stochastic thermodynamics and check the form of the nonequilibrium work relations, including the Jarzynski equality and the Crooks work relations.

\subsection{Revising the Definitions of Work and Heat}
In this subsection, we choose to retain the form of microscopic reversibility~(\ref{eq:microreversibility}) and revise the definitions of work and heat in the current framework of stochastic thermodynamics to be Eqs.~(\ref{eq:undermrratiowork}) and~(\ref{eq:undermrratioheat}), respectively. In this situation, microscopic reversibility should be reformulated as
\begin{equation}  \frac{\rho_{+}[\mathbf{x},\mathbf{p}| x(0),p(0)]}{\rho_{-}[\bar{\mathbf{x}},\bar{\mathbf{p}}| \bar{x}(0),\bar{p}(0)]} = e^{- q'[\mathbf{x},\mathbf{p}]/T} . \label{eq:microreversibilityref}\end{equation}
We keep the form of microscopic reversibility~(\ref{eq:microreversibilityref}) for the following reasons. First, starting from microscopic reversibility and choosing proper initial and final distributions, we can derive most of the fluctuation theorems and nonequilibrium work relations, such as the Jarzynski equality~\cite{Crooks1998}, the Crooks work relations~\cite{Crooks1999PRE,Crooks2000PRE}, and the total entropy production fluctuation theorems~\cite{Seifert2005}. In view of the vital bond role, many researchers believe that microscopic reversibility is fundamental in the framework of stochastic thermodynamics. Second, combining microscopic reversibility~(\ref{eq:microreversibilityref}) with Eq.~(\ref{eq:undertotalentflu}), we can naturally follow the statement of Seifert~\cite{Seifert2005} and separate the total entropy production into two contributions: the entropy variation of the stochastic system itself
\begin{equation}    \Delta s   \equiv -\ln \rho_{+}(x(\tau),p(\tau),\tau)-\left(-\ln \rho_{+}(x(0),p(0),0)\right) \label{eq:undersystementropy}\end{equation}
and the entropy variation of the thermal reservoir (or the medium)
\begin{equation}    \Delta s_{\mathrm{m}}   \equiv -\frac{q'}{T} = \ln \frac{\rho_{+}[\mathbf{x},\mathbf{p}| x(0),p(0)]}{\rho_{-}[\bar{\mathbf{x}},\bar{\mathbf{p}}| \bar{x}(0),\bar{p}(0)]} \label{eq:mediumentropy}\end{equation}
with $\Delta s_{\mathrm{tot}}=\Delta s + \Delta s_{\mathrm{m}}$.

Herein, we set the initial and the final distributions to be stationary distributions. The expression of the stationary distribution can be easily derived from the Fokker-Planck equation~(\ref{eq:underdampedFP}) as
\begin{equation}\rho_{\mathrm{st}}(x,p,\lambda_{1},\lambda_{2}) = \mathrm{e}^{ \frac{1}{T}\left [ F(\lambda_{1},\lambda_{2})- \frac{p^{2}}{2}-U(x,\lambda_{1},\lambda_{2}) \right]},\label{eq:underequilibrium}\end{equation}
where
\begin{equation}F(\lambda_{1},\lambda_{2}) \equiv -T\ln \left[\iint \mathrm{e}^{ -\frac{1}{T}\left (  \frac{p^{2}}{2}+U \right)} \mathrm{d}x \mathrm{d}p \right]\label{eq:freeenergy}\end{equation}
represents the normalization factor. The stationary distribution~(\ref{eq:underequilibrium}) possesses the form of canonical distribution. However, combining the stationary distribution~(\ref{eq:underequilibrium}) with the short-time conditional probability~(\ref{eq:underdampedfstp}) and~(\ref{eq:underdampedrstp}), we can verify that the principle of detailed balance~\cite{Gardiner1985} is broken
\begin{widetext}
\begin{eqnarray}&&\rho_{\mathrm{st}}(x(t),p(t),\lambda_{1},\lambda_{2}) \rho_{+}[x(t+\mathrm{d}t),p(t+\mathrm{d}t)| x(t),p(t)] \nonumber \\ && \ne \rho_{\mathrm{st}}(\bar{x}(\bar{t}),\bar{p}(\bar{t}),\bar{\lambda}_{1},\bar{\lambda}_{2}) \rho_{-}[\bar{x}(\bar{t}+\mathrm{d}\bar{t}),\bar{p}(\bar{t}+\mathrm{d}\bar{t})| \bar{x}(\bar{t}),\bar{p}(\bar{t})]. \label{eq:detailedsymmetry}\end{eqnarray}
\end{widetext}
This is different from systems with the magnetic field as a controlling parameter in which the principle of detailed balance is satisfied (see Appendix~\ref{sec:magneticsl} for details). Consequently the concept of equilibrium cannot be applied to systems with odd controlling parameters if the potential satisfies $U^{\mathrm{r}}(x,\lambda_{1},\lambda_{2}) \ne U(x,\lambda_{1},\lambda_{2})$. The breaking of detailed balance is usually caused by nonequilibrium constraints such as non-conservative force~\cite{Spinney2012PRL,Spinney2012PRE,Ford2012PRE,Lee2013,Yeo2016}. Here we find that odd controlling parameters can also cause ``nonequilibrium'' behaviors through breaking the time-reversal invariance of the system Hamiltonian, i.e., $U^{\mathrm{r}}(x,\lambda_{1},\lambda_{2}) \ne U(x,\lambda_{1},\lambda_{2})$. Without causing ambiguity, we refer to the traditional nonequilibrium thermodynamics and still call the stationary distribution~(\ref{eq:underequilibrium}) the steady-state distribution.

By combining microscopic reversibility~(\ref{eq:microreversibilityref}) with the form of the steady-state distribution~(\ref{eq:underequilibrium}), we can obtain the relation
\begin{equation}  \frac{\rho_{+}[\mathbf{x},\mathbf{p}]}{\rho_{-}[\bar{\mathbf{x}},\bar{\mathbf{p}}]}  = e^{(w'-\Delta F)/T},          \label{eq:ratiotpMR}\end{equation}
which leads to three nonequilibrium work relations:
\begin{equation}  \left \langle    \mathrm{e}^{- w'/T}         \right  \rangle_{+} =  \mathrm{e}^{- \Delta F/T},         \label{eq:JarzyReMR}\end{equation}

\begin{equation}   \frac{\rho_{+}(   w' )}{\rho_{-}(-w')} = \mathrm{e}^{\left( w'-\Delta F \right)/T}, \label{eq:DetailCrooksMR}\end{equation}
and
\begin{equation}  \left \langle    \mathcal{O}         \right \rangle_{+} = \left \langle    \bar{\mathcal{O} }  \mathrm{e}^{-\left(w'-\Delta F\right)/T}      \right \rangle_{-}. \label{eq:IntegralCrooksMR}\end{equation}
$\Delta F \equiv F(\lambda_{1}(\tau),\lambda_{2}(\tau))-F(\lambda_{1}(0),\lambda_{2}(0))$ represents the ``steady-state'' free energy difference. $\mathcal{O}$ is a functional of the forward trajectory, while $\bar{\mathcal{O} }$ is the corresponding functional of the time-reversed dynamical trajectory. We have assumed that the functional satisfies $\mathcal{O}[\mathbf{x},\mathbf{p}]=\bar{\mathcal{O} }[\bar{\mathbf{x}},\bar{\mathbf{p}}]$. $\langle \cdots \rangle_{+}$ represents the ensemble average over trajectories stemming from the initial steady state in the forward driving process. $\langle \cdots \rangle_{-}$ represents the ensemble average over trajectories stemming from the initial steady state in the time-reversed driving process. These three nonequilibrium work relations~(\ref{eq:JarzyReMR}),~(\ref{eq:DetailCrooksMR}) and~(\ref{eq:IntegralCrooksMR}) maintain the respective forms of the Jarzynski equality~\cite{Jarzynski1997}, the detailed Crooks work relation~\cite{Crooks1999PRE}, and the integral Crooks work relation~\cite{Crooks2000PRE}, while the stochastic work expression in these relations is replaced by the revised one~(\ref{eq:undermrratiowork}).

Referring to the steady-state thermodynamics~\cite{Oono1998,Hatano2001}, we can associate part of the total stochastic heat with the trajectory entropy $\Delta s_{\mathrm{A}}$
\begin{eqnarray}  q_{\mathrm{p}}  && =   -T \left( \Delta s_{\mathrm{A}} - \Delta s\right) \nonumber \\ &&= \int \left(  \frac{\mathrm{d} p}{\mathrm{d}t} +   \frac{\partial U}{\partial x}      \right ) \circ \mathrm{d} x  . \label{eq:excessheat}\end{eqnarray}
Note that the expression of partial heat~(\ref{eq:excessheat}) is same as the definition of heat in the current framework of stochastic thermodynamics~(\ref{eq:underdampedworkheat2}).

Considering transitions between steady states and substituting Eq.~(\ref{eq:excessheat}) into relation~(\ref{eq:underAdrivingentropy}), we can obtain the relation
\begin{equation}\frac{\rho_{+}[\mathbf{x},\mathbf{p}]}{\rho_{\mathrm{A}}[\hat{\mathbf{x}},\hat{\mathbf{p}}]}  = e^{(w_{\mathrm{p}}-\Delta F)/T} ,        \label{eq:ratiotpMRA}  \end{equation}
where we have defined partial work as
\begin{eqnarray}  w_{\mathrm{p}}  && \equiv  \Delta e - q_{\mathrm{p}}  \nonumber \\ &&= \int \frac{\partial U(x,\lambda_{1},\lambda_{2})}{\partial \lambda_{1}} \mathrm{d} \lambda_{1}+ \int \frac{\partial U(x,\lambda_{1},\lambda_{2})}{\partial \lambda_{2}}  \mathrm{d} \lambda_{2}.  \label{eq:excesswork}\end{eqnarray}
According to relation~(\ref{eq:ratiotpMRA}), we can also derive three nonequilibrium work relations:
\begin{equation}  \left \langle    \mathrm{e}^{- w_{\mathrm{p}}/T}         \right  \rangle_{+} =  \mathrm{e}^{- \Delta F/T} ,         \label{eq:JarzyReIA}\end{equation}

\begin{equation}   \frac{\rho_{+}(   w_{\mathrm{p}} )}{\rho_{\mathrm{A}}(-w_{\mathrm{p}})} = \mathrm{e}^{\left( w_{\mathrm{p}}-\Delta F   \right)/T}, \label{eq:DetailCrooksIA}\end{equation}
and
\begin{equation}  \left \langle    \mathcal{O}         \right \rangle_{+} = \left \langle    \hat{\mathcal{O} }  \mathrm{e}^{-\left(w_{\mathrm{p}}-\Delta F\right)/T}      \right \rangle_{\mathrm{A}}. \label{eq:IntegralCrooksIA}\end{equation}
$\langle \cdots \rangle_{\mathrm{A}}$ represents the ensemble average over trajectories stemming from the initial steady state in the $\mathrm{A}$ adjoint driving process. Comparing with the previous three nonequilibrium work relations~(\ref{eq:JarzyReMR}),~(\ref{eq:DetailCrooksMR}), and~(\ref{eq:IntegralCrooksMR}), we can find that the above three nonequilibrium work relations~(\ref{eq:JarzyReIA}),~(\ref{eq:DetailCrooksIA}), and~(\ref{eq:IntegralCrooksIA}) also retain the respective forms of the Jarzynski equality~\cite{Jarzynski1997}, the detailed Crooks work relation~\cite{Crooks1999PRE}, and the integral Crooks work relation~\cite{Crooks2000PRE}. However, it is worth noting that the partial work $w_{\mathrm{p}}$ we consider here is just part of the total work $w'$. In addition, the time-reversed driving process is replaced by the $\mathrm{A}$ adjoint driving process.

\subsection{Revising Microscopic Reversibility}
In this subsection, we instead keep the form of the definitions of work~(\ref{eq:underdampedworkheat1}) and heat~(\ref{eq:underdampedworkheat2}) in the current framework of stochastic thermodynamics and instead revise microscopic reversibility. We retain the form of the definitions of work and heat for the following reasons. First, the form of the definitions of work~(\ref{eq:underdampedworkheat1}) and heat~(\ref{eq:underdampedworkheat2}) seem a natural means of extension from systems containing one kind of controlling parameter to systems containing two types of controlling parameters. Second, the form of the definitions of work~(\ref{eq:underdampedworkheat1}) and heat~(\ref{eq:underdampedworkheat2}) coincide with the thermodynamic interpretation of the Langevin dynamics given by Sekimoto~\cite{Sekimoto2010}. This makes the definitions of work~(\ref{eq:underdampedworkheat1}) and heat~(\ref{eq:underdampedworkheat2}) physically easier to understand than the alternative definitions of work~(\ref{eq:undermrratiowork}) and heat~(\ref{eq:undermrratioheat}).

According to the definition of heat~(\ref{eq:underdampedworkheat2}), the microscopically reversible condition should be revised to be
\begin{equation}  \frac{\rho_{+}[\mathbf{x},\mathbf{p}| x(0),p(0)]}{\rho_{-}[\bar{\mathbf{x}},\bar{\mathbf{p}}| \bar{x}(0),\bar{p}(0)]} = e^{- q/T +I }  \label{eq:microreversibility1}\end{equation}
with
\begin{widetext}
\begin{eqnarray} I[\mathbf{x},\mathbf{p}] \equiv && \ln \frac{\rho_{\mathrm{A}}[\hat{\mathbf{x}},\hat{\mathbf{p}}]}{\rho_{-}[\bar{\mathbf{x}},\bar{\mathbf{p}}]}
\nonumber \\ =&&\frac{1}{2T}\int \left( \frac{\partial U}{\partial x} -\frac{\partial U^{\mathrm{r}}}{\partial x} \right) \circ \mathrm{d}x - \frac{1}{2\gamma T} \int \left[  \frac{\mathrm{d} p}{\mathrm{d}t} + \frac{1}{2} \left(  \frac{\partial U}{\partial x} + \frac{\partial U^{\mathrm{r}}}{\partial x}    \right)  \right]   \left(  \frac{\partial U}{\partial x} - \frac{\partial U^{\mathrm{r}}}{\partial x}    \right)\mathrm{d}t  \label{eq:extraI}\end{eqnarray}
\end{widetext}
representing the correction term caused by the inclusion of odd controlling parameters. This correction term is related with the trajectory entropy by the relation
\begin{equation}   I = \Delta s_{\mathrm{tot}} - \Delta s_{\mathrm{A}} =  \Delta s_{\mathrm{B}} + \Delta s_{\mathrm{C}}  .   \label{eq:Irelatedte}\end{equation}
In the particular situation $U^{\mathrm{r}}(x,\lambda_{1},\lambda_{2}) = U(x,\lambda_{1},\lambda_{2})$, this correction term vanishes. Since this correction term do not obey an integral fluctuation theorem, we cannot give any bounds on the sign of its mean.

Considering transitions between stationary states and taking revised microscopic reversibility~(\ref{eq:microreversibility1}) into account, we can obtain the relation
\begin{equation}  \frac{\rho_{+}[\mathbf{x},\mathbf{p}]}{\rho_{-}[\bar{\mathbf{x}},\bar{\mathbf{p}}]}  = e^{(w-\Delta F)/T+I},          \label{eq:ratiotp}\end{equation}
which leads to three nonequilibrium work relations:
\begin{equation}  \left \langle    \mathrm{e}^{- w/T-I}         \right  \rangle_{+} =  \mathrm{e}^{- \Delta F/T} ,         \label{eq:JarzyReI}\end{equation}

\begin{equation}   \frac{\rho_{+}(   w )}{\rho_{-}(-w)} = \mathrm{e}^{\left( w-\Delta F \right)/T+I}, \label{eq:DetailCrooksI}\end{equation}
and
\begin{equation}  \left \langle    \mathcal{O}         \right \rangle_{+} = \left \langle    \bar{\mathcal{O} }  \mathrm{e}^{-\left(w-\Delta F\right)/T-I}      \right \rangle_{-}. \label{eq:IntegralCrooksI}\end{equation}
This indicates that for systems with odd controlling parameters, if we keep the form of the definitions of work~(\ref{eq:underdampedworkheat1}) and heat~(\ref{eq:underdampedworkheat2}) in the current framework of stochastic thermodynamics, a correction term $I$ needs to be added to revised versions of microscopic reversibility~\cite{Crooks1998}, the Jarzynski equality~\cite{Jarzynski1997}, and the Crooks work relations~\cite{Crooks1999PRE,Crooks2000PRE}.

Note that we can derive a modified second law of thermodynamics
\begin{equation}   \langle w \rangle \ge \Delta F - T \langle I \rangle  \label{eq:modifiedsl}\end{equation}
from Eq.~(\ref{eq:JarzyReI}) by using the Jensen inequality. Here we have set the Boltzmann factor to be unit for simplicity. The average correction term $\langle I \rangle$ can be positive or negative. It looks similar to the mutual information in the feedback-driven process~\cite{Sagawa2010}. However, we are not using any feedback information in the present framework. Such a correction term results from odd controlling parameters. It is meaningful to investigate the form of the protocol and the potential that can keep the average correction term $\langle I \rangle$ to be negative. In this case, equation~(\ref{eq:modifiedsl}) will give a stronger bound for the maximal mean extractable work $-\langle w \rangle$. We leave this issue for future works.

On the other hand, if we start from Eq.~(\ref{eq:underAdrivingentropy}) and consider transitions between stationary states, we can derive the relation
\begin{equation}  \frac{\rho_{+}[\mathbf{x},\mathbf{p}]}{\rho_{\mathrm{A}}[\hat{\mathbf{x}},\hat{\mathbf{p}}]}  = e^{(w-\Delta F)/T}.          \label{eq:ratiotpA}\end{equation}
Comparing with Eq.~(\ref{eq:ratiotp}), we can find that the time-reversed driving process is replaced by the $\mathrm{A}$ adjoint driving process in Eq.~(\ref{eq:ratiotpA}), but the additional quantity $I[\mathbf{x},\mathbf{p}]$ is successfully eliminated. Three different nonequilibrium work relations can also be derived from Eq.~(\ref{eq:ratiotpA}):
\begin{equation}  \left \langle    \mathrm{e}^{- w/T}         \right  \rangle_{+} =  \mathrm{e}^{- \Delta F/T} ,         \label{eq:JarzyReA}\end{equation}

\begin{equation}   \frac{\rho_{+}(   w )}{\rho_{\mathrm{A}}(-w)} = \mathrm{e}^{\left( w-\Delta F \right)/T}, \label{eq:DetailCrooksA}\end{equation}
and
\begin{equation}  \left \langle    \mathcal{O}         \right \rangle_{+} = \left \langle    \hat{\mathcal{O} }  \mathrm{e}^{-\left(w-\Delta F\right)/T}      \right \rangle_{\mathrm{A}}. \label{eq:IntegralCrooksA}\end{equation}
These three nonequilibrium work relations~(\ref{eq:JarzyReA}),~(\ref{eq:DetailCrooksA}), and~(\ref{eq:IntegralCrooksA}) mathematically recover the Jarzynski equality~\cite{Jarzynski1997}, the detailed Crooks work relation~\cite{Crooks1999PRE}, and the integral Crooks work relation~\cite{Crooks2000PRE}, respectively. However it is worth noting that the time-reversed driving process in the traditional detailed and integral Crooks work relations~\cite{Crooks1999PRE,Crooks2000PRE} is replaced by the $\mathrm{A}$ adjoint driving process in Eqs.~(\ref{eq:DetailCrooksA}) and~(\ref{eq:IntegralCrooksA}).

\section{Example: Shortcuts to Isothermality\label{examplesti}}
To explain our results explicitly, we consider the strategy of shortcuts to isothermality~\cite{Geng2017} as an example. In conventional thermodynamics, it is widely believed that the realization of an isothermal process needs to quasi-statically drive controlling parameters. The strategy of shortcuts to isothermality is designed to realize a finite-rate isothermal transition between two equilibrium states with the same temperature~\cite{Geng2017}. Within the above framework of stochastic thermodynamics for systems with odd controlling parameters, we can give a deeper understanding of shortcuts to isothermality as well as the Jarzynski-like equality derived from it.

We first briefly introduce the strategy of shortcuts to isothermality; please refer to~\cite{Geng2017} for further details. Within the framework of shortcuts to isothermality, an auxiliary potential $U_{1}(x,t)$ is introduced to the system of interest with the Hamiltonian $H_{0}(x,p,\Lambda)=p^{2}/2+U_{0}(x,\Lambda)$. Herein, $\Lambda \equiv \Lambda(t)$ is the controlling parameter. This auxiliary potential is required to escort the evolution of the system so that the system distribution is always in the instantaneous equilibrium distribution of the original Hamiltonian $H_{0}(x,p,\Lambda(t))$:
\begin{equation} \rho(x,p,t)=\rho_{\mathrm{ieq}}(x,p,\Lambda) \equiv \mathrm{e}^{ \left [ F_{0}(\Lambda)- H_{0}(x,p,\Lambda)\right]/T}\label{eq:STequilibrium}\end{equation}
with
\begin{equation}F_{0}(\Lambda) \equiv -T\ln \left[\iint \mathrm{e}^{- H_{0}(x,p,\Lambda)/T}\mathrm{d}x \mathrm{d}p\right]\label{eq:STfreeenergy}\end{equation}
representing the free energy of the original system in equilibrium. To this end, the auxiliary potential is demonstrated to have the structure
\begin{equation}     U_{1}(x,t)=\dot{\Lambda}f(x,\Lambda),                                 \label{eq:STauxpot}\end{equation}
where function $f(x,\Lambda)$ can be determined according to the method in~\cite{Geng2017} and $\dot{\Lambda} \equiv \mathrm{d} \Lambda(t) /\mathrm{d} t$. By imposing boundary conditions
\begin{equation}\dot{\Lambda}(0) = \dot{\Lambda}(\tau) = 0,\label{eq:STboundary}\end{equation}
we can make the auxiliary potential vanish at the beginning $t=0$ and end $t=\tau$ of the driving process, which indicates that the distribution functions at the two endpoints of the driving process become equilibrium distributions. This is the main idea of shortcuts to isothermality.

As the time derivative of the controlling parameter changes its sign during time-reversal operation, i.e.,
\begin{equation}     \frac{\mathrm{d} \bar{\Lambda}(\bar{t})}{\mathrm{d}\bar{ t}} = -  \frac{\mathrm{d} \Lambda(t)}{ \mathrm{d} t   } ,  \label{eq:STtimerever}\end{equation}
we can treat it as the odd controlling parameter and map shortcuts to isothermality into the stochastic thermodynamic framework for systems with odd controlling parameters. The mapping relations read
\begin{equation}   \lambda_{1}(t) = \Lambda(t), ~~~ \lambda_{2}(t)=\dot{\Lambda}(t),   \label{eq:STmappingre1}\end{equation}
and
\begin{eqnarray}    U(x,\lambda_{1},\lambda_{2}) && = U_{0}(x,\lambda_{1})+ U_{1}(x,\lambda_{1},\lambda_{2}) \nonumber \\ &&= U_{0}(x,\lambda_{1}) +  \lambda_{2}f(x,\lambda_{1})  .   \label{eq:STmappingre2}\end{eqnarray}
Therefore, during the driving process of shortcuts to isothermality, the system cannot reach the equilibrium state because of the existence of the odd controlling parameter $\lambda_{2}=\dot{\Lambda}(t)$. Only at the two endpoints of the driving process, the odd controlling parameter vanishes, allowing the system to evolve to equilibrium state.

Three nonequilibrium work relations were derived under the framework of shortcuts to isothermality in~\cite{Geng2017}. Among the three nonequilibrium work relations, the third relation, a Jarzynski-like equality, is closely related to the choice of the definition of work. In the following, we discuss the relationship between this Jarzynski-like equality and the definition of work.

In~\cite{Geng2017}, the definition of work maintains the form~(\ref{eq:underdampedworkheat1}). In this situation, the integral Crooks work relation is revised to be Eqs.~(\ref{eq:IntegralCrooksI}) or~(\ref{eq:IntegralCrooksA}).

Starting from Eq.~(\ref{eq:IntegralCrooksA}) and assuming that $\mathcal{O}=\delta(x-x(\tau))\delta(p-p(\tau))$, we can obtain the relation
\begin{eqnarray}  \rho_{+}(x,p,\tau)  && =  \langle  \delta(x-x(\tau))\delta(p-p(\tau)) \rangle_{+} \nonumber \\  && = \langle  \delta(x-\hat{x}(0))\delta(p-\hat{p}(0)) \mathrm{e}^{-(w-\Delta F)/T} \rangle_{\mathrm{A}} \nonumber \\  && =  \rho_{\mathrm{st}}(x,p) \langle \mathrm{e}^{-(w-\Delta F)/T} \rangle_{\{x,p\},A} .\label{eq:STJarlikeII}\end{eqnarray}
Here $\langle \cdots \rangle_{\{x,p\},A}$ represents the ensemble average over all trajectories starting from a fixed state $\{x,p\}$ in the $\mathrm{A}$ adjoint driving process. $\rho_{+}(x,p,\tau)$ and $\rho_{\mathrm{st}}(x,p)$ respectively represent the final state in the forward driving process and the corresponding steady state when the controlling parameter is fixed. $\Delta F$ represents the free energy difference of the system with the auxiliary potential~(\ref{eq:STauxpot}). Applying the strategy of shortcuts to isothermality, we can evolve the system from an equilibrium state to another one at the same temperature, i.e., $\rho_{+}(x,p,\tau)=\rho_{\mathrm{st}}(x,p)=\rho_{\mathrm{eq}}(x,p)$ in Eq.~(\ref{eq:STJarlikeII}). Therefore, we can derive the equality
\begin{equation}  \langle \mathrm{e}^{- (w-\Delta F)/T} \rangle_{\{x,p\},A} = 1.  \label{eq:STthirdeq}\end{equation}

In the forward driving process, the system Hamiltonian is
\begin{eqnarray}  H(x,p,t)= H_{0}(x,p,\lambda_{1}(t)) + U_{1}(x,\lambda_{1}(t),\lambda_{2}(t))   .   \label{eq:STtotalH}\end{eqnarray}
Considering the mapping relation between the $\mathrm{A}$ adjoint driving process and the forward driving process~(\ref{eq:underAmaprela}), we can derive the system Hamiltonian in the $\mathrm{A}$ adjoint driving process as
\begin{eqnarray}  H(\hat{x},\hat{p},\hat{t})= H_{0}(\hat{x},\hat{p},\hat{\lambda}_{1}(\hat{t})) + U_{1}(\hat{x},\hat{\lambda}_{1}(\hat{t}),\hat{\lambda}_{2}(\hat{t}))   .   \label{eq:STtotalHad}\end{eqnarray}
Note that the system Hamiltonian in the $\mathrm{A}$ adjoint driving process~(\ref{eq:STtotalHad}) has the same structure as the one in the forward driving process~(\ref{eq:STtotalH}). This means that in the $\mathrm{A}$ adjoint driving process, the strategy of shortcuts to isothermality can also drive the system from an equilibrium state to another one at the same temperature. According to this property of shortcuts to isothermality, we can omit the subscript `$\mathrm{A}$' in Eq.~(\ref{eq:STthirdeq}) and simplify it to be
\begin{equation}  \langle \mathrm{e}^{- (w-\Delta F)/T} \rangle_{\{x,p\}} = 1.  \label{eq:STthirdeq2}\end{equation}
This equation is a Jarzynski-like equality and implies that one can estimate $\Delta F$ by taking the exponential average of the work $w$ over trajectories that start from an arbitrary fixed state $\{x,p\}$ and then evolve under the strategy of shortcuts to isothermality.

Equation~(\ref{eq:STthirdeq2}) is derived from the revised integral Crooks work relation~(\ref{eq:IntegralCrooksA}). If we instead start from Eq.~(\ref{eq:IntegralCrooksI}), we can also derive a Jarzynski-like equality
\begin{equation}  \langle \mathrm{e}^{- (w-\Delta F)/T-I} \rangle_{\{x,p\},-} = 1.  \label{eq:STthirdeqIII}\end{equation}
In the time-reversed driving process, the system Hamiltonian is expressed as
\begin{eqnarray}  H(\bar{x},\bar{p},\bar{t}) && = H_{0}(\bar{x},\bar{p},\bar{\lambda}_{1}(\bar{t})) + U_{1}(\bar{x},\bar{\lambda}_{1}(\bar{t}),-\bar{\lambda}_{2}(\bar{t}))  \nonumber \\  && = H_{0}(\bar{x},\bar{p},\bar{\lambda}_{1}(\bar{t})) + U^{r}_{1}(\bar{x},\bar{\lambda}_{1}(\bar{t}),\bar{\lambda}_{2}(\bar{t})) .   \label{eq:STtotalHadIII}\end{eqnarray}
Compared with the system Hamiltonian in the forward driving process~(\ref{eq:STtotalH}), we find that in the time-reversed driving process, the strategy of shortcuts to isothermality cannot drive the system from an equilibrium state to another one at the same temperature. Therefore, the subscript `$-$' in Eq.~(\ref{eq:STthirdeqIII}) cannot be omitted directly.

The Jarzynski-like equalities~(\ref{eq:STthirdeq2}) and~(\ref{eq:STthirdeqIII}) are all derived under the choice of keeping the form of the definition of stochastic work~(\ref{eq:underdampedworkheat1}) unchanged. If we instead choose to retain the form of the microscopically reversible condition~(\ref{eq:microreversibilityref}) and revise the definition of work to be~(\ref{eq:undermrratiowork}), the integral Crooks work relation is revised to be Eqs.~(\ref{eq:IntegralCrooksMR}) or~(\ref{eq:IntegralCrooksIA}). Starting from these two relations, we can similarly derive two Jarzynski-like equalities
\begin{equation}  \langle \mathrm{e}^{- (w'-\Delta F)/T} \rangle_{\{x,p\},-} = 1  \label{eq:STthirdeqIIII}\end{equation}
and
\begin{equation}  \langle \mathrm{e}^{- (w_{\mathrm{p}}-\Delta F)/T} \rangle_{\{x,p\},A} = 1.  \label{eq:STthirdeqIIIII}\end{equation}
As the system Hamiltonian in the $\mathrm{A}$ adjoint driving process~(\ref{eq:STtotalHad}) has the same structure as the one in the forward driving process~(\ref{eq:STtotalH}), the subscript `$\mathrm{A}$' in Eq.~(\ref{eq:STthirdeqIIIII}) can be omitted, i.e.,
\begin{equation}  \langle \mathrm{e}^{- (w_{\mathrm{p}}-\Delta F)/T} \rangle_{\{x,p\}} = 1.  \label{eq:STthirdeqIIIIIA}\end{equation}
However, the subscript `$-$' in Eq.~(\ref{eq:STthirdeqIIII}) cannot be omitted directly because of the different structures between the system Hamiltonian~(\ref{eq:STtotalH}) and~(\ref{eq:STtotalHadIII}).

\section{Conclusion and Discussion\label{disclu}}

In this paper, we constructed a stochastic thermodynamic framework for systems with odd controlling parameters. With the inclusion of odd controlling parameters, we find that the definition of heat is incompatible with the microscopically reversible condition when applying the current framework of stochastic thermodynamics to systems with odd controlling parameters. Such a contradiction also induces a revision to the fluctuation theorems and the nonequilibrium work relations. By introducing adjoint dynamics, the total entropy production $\Delta s_{\mathrm{tot}}$ can be separated into three parts $\Delta s_{\mathrm{A}}$, $\Delta s_{\mathrm{B}}$, and $\Delta s_{\mathrm{C}}$. The total entropy production $\Delta s_{\mathrm{tot}}$, the trajectory entropy $\Delta s_{\mathrm{A}}$, and the trajectory entropy $\Delta s_{\mathrm{B}}$ are respectively shown to satisfy the integral fluctuation theorem. By revising the definitions of work and heat and the expression of microscopic reversibility, we obtained two sets of modified nonequilibrium work relations, including the Jarzynski equality, the detailed Crooks work relation, and the integral Crooks work relation. We considered the strategy of shortcuts to isothermality as an example. According to the two different choices of the definition of work, we gave different explanations for the Jarzynski-like equality derived from shortcuts to isothermality.

In the above framework, we have assumed that odd controlling parameters are only contained in the potential. It is straightforward to extend our framework to more general cases in which odd controlling parameters can appear in the whole Hamiltonian. A familiar example is when systems take the magnetic field as a controlling parameter, where the magnetic field appears in the kinetic term. The magnetic field as an odd controlling parameter has already been widely investigated in the field of nonequilibrium thermodynamics~\cite{Seifert2012,Saha2008,Baiesi2011}. Since the Hamiltonian for systems with the magnetic field as the only odd controlling parameter is always time-reversal invariant, the correction term $I$ to the Jarzynski equality~(\ref{eq:JarzyReI}) or the second law of thermodynamics~(\ref{eq:modifiedsl}) will vanish in this situation (see Appendix~\ref{sec:magneticsl} for details). In contrast, the applied potential or the Hamiltonian in our framework is not necessarily an even function of odd controlling parameters, for example, shortcuts to isothermality~\cite{Geng2017}, active Ornstein-Uhlenbeck processes~\cite{Mandal2017}, isothermal-isobaric molecular dynamics~\cite{Evans1990}, and other controlling theories~\cite{Adib2005,Vaikuntanathan2008,Ballard2009}. In this situation, the odd controlling parameters break the time-reversal invariance of the Hamiltonian and lead to a series of novel fluctuation relations in this paper.

In section~\ref{nonwore}, we have proposed two schemes to solve the contradiction between the definition of heat and microscopic reversibility. Heat or work along the stochastic trajectory may need to be measured experimentally for small systems with odd controlling parameters to determine which scheme is more suitable. Differential fluctuation theorem has been experimentally verified in both underdamped and overdamped situations for systems containing only even controlling parameters~\cite{Hoang2018}. Similar experiment may be conducted for small systems with odd controlling parameters to answer the question above. We prefer to choose the scheme of keeping the definitions of work and heat and revising the expression of microscopic reversibility because it is compatible with the thermodynamic interpretation of the Langevin dynamics. In this scheme, we can express heat as $[-\gamma p + \xi(t)]\circ \mathrm{d} x$ through substituting the Langevin equation~(\ref{eq:underdampedLE}) into Eq.~(\ref{eq:underdampedworkheat2}). From the standpoint of the Brownian motion, the energy $[-\gamma p + \xi(t)]\circ \mathrm{d} x$ can be naturally interpreted as work done on the system by the thermal reservoir~\cite{Sekimoto2010}. This makes the definitions of work~(\ref{eq:underdampedworkheat1}) and heat~(\ref{eq:underdampedworkheat2}) physically easier to understand than the alternative definitions of work~(\ref{eq:undermrratiowork}) and heat~(\ref{eq:undermrratioheat}).

We have discussed the stochastic thermodynamic framework in the underdamped situation. It is straightforward to extend this framework to the overdamped situation. In addition, our discussion has assumed that the system does not contain a momentum-dependent driving force~\cite{Kim2004,Kim2007,Schweitzer2002}. Such a force has induced rich phenomena for small systems~\cite{Spinney2012PRL,Lee2013,Spinney2012PRE,Ford2012PRE,Yeo2016,Kim2004,Kim2007}. The investigation of small systems with odd controlling parameters will be an interesting topic for future work, as will systems possessing a momentum-dependent driving force.

\emph{Acknowledgement.}--The authors are grateful to financial support from the
National Natural Science Foundation of China (Grant NOs. 11675017 and 11735005) and the Fundamental Research Funds for the Central Universities (NO. 2017EYT24).

\appendix

\section{Detailed derivations of short-time conditional probability\label{sec:shorttime}}
In this section, we derive the short-time conditional probability for systems with odd controlling parameters.
The Fokker-Planck equation~(\ref{eq:underdampedFP}) can be abbreviated as
\begin{equation}\frac{\partial \rho}{\partial t} =  \frac{\partial }{\partial x} (D_{1} \rho) + \frac{\partial }{\partial p} (D_{2} \rho) + \frac{\partial^{2} }{\partial p^{2}} (D_{3} \rho) ,\label{app:underfkeq}\end{equation}
where
\begin{equation}  D_{1} =  -p, ~~~D_{2} =  \gamma p + \frac{\partial U}{\partial x}, ~~~ D_{3} =\gamma T.\label{app:undercoeff}\end{equation}
After discretization, we can express the left side of Eq.~(\ref{app:underfkeq}) as
\begin{eqnarray} \mathrm{left} =  &&\frac{ \rho(x',p',t+ \Delta t | x,p,t) - \rho(x',p',t  | x,p,t)}{\Delta t} \nonumber \\= &&\frac{ \rho(x',p',t+ \Delta t | x,p,t) - \delta (x'-x) \delta (p'-p)}{\Delta t}
 ,\label{app:underleft}\end{eqnarray}
where we have included abbreviations $x' \equiv x(t+ \Delta t)$, $p' \equiv p(t+ \Delta t)$, $x \equiv x(t)$, and $p \equiv p(t)$. The right side follows
\begin{widetext}
\begin{eqnarray} \mathrm{right} = \left(   \frac{\partial }{\partial x'} D_{1}(x',p') + \frac{\partial }{\partial p'} D_{2}(x',p') +       \frac{\partial^{2} }{\partial {p'}^{2}} D_{3}(x',p')\right)\delta(x'-x)\delta(p'-p).    \label{app:underright}\end{eqnarray}
Combining Eqs.~(\ref{app:underleft}) and~(\ref{app:underright}), we can obtain the conditional probability as
\begin{eqnarray}   \rho(x',p',t+ \Delta t | x,p,t)= \left(  1+ \Delta t \frac{\partial }{\partial x'} D_{1}(x',p') +\Delta t \frac{\partial }{\partial p'} D_{2}(x',p')        +      \Delta t \frac{\partial^{2} }{\partial {p'}^{2}} D_{3}(x',p')\right)\delta(x'-x)\delta(p'-p).    \label{app:underconditional}\end{eqnarray}
In the short-time limit $\Delta t \to \mathrm{d} t $, we can then derive the short-time conditional probability
\begin{eqnarray}\rho(x',p',t+\mathrm{d} t| x,p,t)  =&& \left(  1+ \mathrm{d} t  \frac{\partial }{\partial x'} D_{1}(x',p') +\mathrm{d} t  \frac{\partial }{\partial p'} D_{2}(x',p')       +      \mathrm{d} t  \frac{\partial^{2} }{\partial {p'}^{2}} D_{3}(x',p')\right)\delta(x'-x)\delta(p'-p)   \nonumber \\ =&& \left(  1+ D_{1}(x,p)\mathrm{d} t  \frac{\partial }{\partial x'}  +  D_{2}(x,p)\mathrm{d} t\frac{\partial }{\partial p'}       +      D_{3}(x,p) \mathrm{d} t \frac{\partial^{2} }{\partial {p'}^{2}} \right)\delta(x'-x)\delta(p'-p)  \nonumber \\ = &&\frac{1}{(2\pi)^{2}} \left(  1+ D_{1}\mathrm{d} t \frac{\partial }{\partial x'}  + D_{2}\mathrm{d} t \frac{\partial }{\partial p'}       +      D_{3}\mathrm{d} t \frac{\partial^{2} }{\partial {p'}^{2}} \right) \iint e^{iu(x'-x)+iv(p'-p)} dudv  \nonumber \\ = &&\frac{1}{(2\pi)^{2}} \iint \left(  1+ iu D_{1} \mathrm{d} t   + iv D_{2}\mathrm{d} t  - v^{2}      D_{3}\mathrm{d} t   \right)  e^{iu(x'-x)+iv(p'-p)} dudv
  \nonumber \\ =&& \frac{1}{(2\pi)^{2}} \iint e^{iu(x'-x+ D_{1}\mathrm{d} t )}  e^{- D_{3} v ^{2}\mathrm{d} t +iv(p'-p+ D_{2}\mathrm{d} t )} dudv
  \nonumber \\ =&& \delta (x'-x+ D_{1}\mathrm{d} t )\frac{1}{\sqrt{4 \pi D_{3} \mathrm{d} t }} e^{-\frac{(p'-p+D_{2} \mathrm{d} t  )^{2}}{4D_{3}\mathrm{d} t }}
 . \label{app:understcp}\end{eqnarray}
 \end{widetext}
We have substituted the Fourier expansion $\delta (x)=1/2 \pi \int \mathrm{e}^{iux}\mathrm{d}u$ in the third line. From the fourth line to the fifth line, we have taken advantage of the first-order approximation for the exponential function. Substituting Eq.~(\ref{app:undercoeff}) into the above equation, we can derive the form of short-time conditional probability~(\ref{eq:underdampedfstp}) found in the main text.

\section{Expressions for $\Delta s_{\mathrm{A}}$, $\Delta s_{\mathrm{B}}$, and $\Delta s_{\mathrm{C}}$\label{sec:expression}}
In this part, we give detailed derivation of expressions for trajectory entropy $\Delta s_{\mathrm{A}}$, $\Delta s_{\mathrm{B}}$, and $\Delta s_{\mathrm{C}}$.
Taking the relation~(\ref{eq:adjointdynamics}) into account, we can derive the short-time conditional probability for $\mathrm{A}$ adjoint driving processes as
\begin{widetext}
\begin{eqnarray}    \rho_{\mathrm{A}}[\hat{x}(\hat{t}+\mathrm{d}\hat{t}),\hat{p}(\hat{t}+\mathrm{d}\hat{t})|\hat{x}(\hat{t}),\hat{p}(\hat{t})] =\frac{ \delta (\mathrm{d}\hat{x} - \hat{p} \mathrm{d}\hat{t}) }{\sqrt{4 \pi \gamma T \mathrm{d}\hat{t}}} \mathrm{exp}{\left[-\frac{\left(\mathrm{d}\hat{p}+\gamma \hat{p} \mathrm{d}\hat{t}- \frac{\partial U(\hat{x},\hat{\lambda}_{1},\hat{\lambda}_{2})}{\partial \hat{x}} \mathrm{d} \hat{t}      \right)^{2}}{4 \gamma T\mathrm{d} \hat{t}}\right]}  .\label{app:underastcp}\end{eqnarray}
\end{widetext}
Substituting the short-time conditional probability~(\ref{eq:underdampedfstp}) and~(\ref{app:underastcp}) into Eq.~(\ref{eq:underAdrivingentropy}), we can express the trajectory entropy $\Delta s_{\mathrm{A}}$ as
\begin{eqnarray}  \Delta s_{\mathrm{A}} = &&  \ln \left(  \frac{\rho_{+}(x(0),p(0),0)}{\rho_{\mathrm{A}}(\hat{x}(0),\hat{p}(0),0)} \right)   +   \ln \left( \frac{\rho_{+}[\mathbf{x},\mathbf{p}| x(0),p(0)]}{\rho_{\mathrm{A}}[\hat{\mathbf{x}},\hat{\mathbf{p}}| \hat{x}(0),\hat{p}(0)]} \right) \nonumber \\ =&&  \ln \left(  \frac{\rho_{+}(x,,p,0)}{\rho_{+}(x,p,\tau)} \right) -\frac{1}{T}\int \left( \frac{\mathrm{d}p}{\mathrm{d}t}+\frac{\partial U}{\partial x} \right)   \circ\mathrm{d} x. \label{eq:underaexpre}\end{eqnarray}
Herein, we have considered the initial distribution of the $\mathrm{A}$ adjoint driving process~(\ref{eq:Ainitial}) and the mapping relation~(\ref{eq:underAmaprela}).

For the $\mathrm{B}$ adjoint driving process, the evolution of the system also follows the adjoint dynamics. Therefore, we can express its short-time conditional probability as
\begin{widetext}
\begin{eqnarray}    \rho_{\mathrm{B}}[\tilde{x}(\tilde{t}+\mathrm{d}\tilde{t}),\tilde{p}(\tilde{t}+\mathrm{d}\tilde{t})|\tilde{x}(\tilde{t}),\tilde{p}(\tilde{t})] =\frac{ \delta (\mathrm{d}\tilde{x} - \tilde{p} \mathrm{d}\tilde{t}) }{\sqrt{4 \pi \gamma T \mathrm{d}\tilde{t}}} \mathrm{exp}{\left[-\frac{\left(\mathrm{d}\tilde{p}+\gamma \tilde{p} \mathrm{d}\tilde{t}- \frac{\partial U(\tilde{x},\tilde{\lambda}_{1},\tilde{\lambda}_{2})}{\partial \tilde{x}}  \mathrm{d} \tilde{t}     \right)^{2}}{4 \gamma T\mathrm{d} \tilde{t}}\right]}  .\label{app:underbstcp}\end{eqnarray}

Substituting Eq.~(\ref{app:underbstcp}) into the definition of $\Delta s_{\mathrm{B}}$~(\ref{eq:underBdrivingentropy}), we can derive that
\begin{eqnarray}  \Delta s_{\mathrm{B}} = &&  \ln \left(  \frac{\rho_{+}(x(0),p(0),0)}{\rho_{\mathrm{B}}(\tilde{x}(0),\tilde{p}(0),0)} \right)   +   \ln \left( \frac{\rho_{+}[\mathbf{x},\mathbf{p}| x(0),p(0)]}{\rho_{\mathrm{B}}[\tilde{\mathbf{x}},\tilde{\mathbf{p}}| \tilde{x}(0),\tilde{p}(0)]} \right) \nonumber \\ = &&  -\frac{1}{2T} \int \left( \frac{\partial U}{\partial x}  -  \frac{\partial U^{\mathrm{r}}}{\partial x}     \right)    \circ \mathrm{d} x-  \frac{1}{2\gamma T}   \int^{\tau}_{0} \left [    \frac{\mathrm{d} p}{\mathrm{d}t}   +  \frac{1}{2} \left( \frac{\partial U}{\partial x}  +  \frac{\partial U^{\mathrm{r}}}{\partial x}     \right)     \right ]   \left( \frac{\partial U}{\partial x}  -  \frac{\partial U^{\mathrm{r}}}{\partial x}     \right) \mathrm{d} t , \label{eq:underbexpre}\end{eqnarray}
\end{widetext}
where the initial distribution of the $\mathrm{B}$ adjoint driving process~(\ref{eq:Binitial}) and the mapping relation~(\ref{eq:underBmaprela}) have been considered.

Subtracting trajectory entropy $\Delta s_{\mathrm{A}}$ and $\Delta s_{\mathrm{B}}$ from the total entropy $\Delta s_{\mathrm{tot}}$~(\ref{eq:undertotalentropy}), we obtain the expression for $\Delta s_{\mathrm{C}}$ as
\begin{eqnarray}  \Delta s_{\mathrm{C}} = && \Delta s_{\mathrm{tot}} -  \Delta s_{\mathrm{A}}  -\Delta s_{\mathrm{B}} \nonumber \\ =&&   \frac{1}{T}\int \left(  \frac{\partial U}{\partial x}  -  \frac{\partial U^{\mathrm{r}}}{\partial x} \right) \circ \mathrm{d} x. \label{eq:undercexpre}\end{eqnarray}

\section{Small systems with the magnetic field as a controlling parameter\label{sec:magneticsl}}
Consider a charged Brownian particle moving in a harmonic potential. A time-dependent magnetic field $B(t)$ is applied in the $z$ direction. The Hamiltonian of the system is
\begin{equation}    H=\frac{1}{2} \left [  \left( p_{x} +\frac{B}{2}y   \right)^{2} +  \left( p_{y} -\frac{B}{2}x   \right)^{2}   \right]  + \frac{k}{2} \left(x^{2}+y^{2} \right ) ,\label{app:chargedhamiltonian}\end{equation}
where $k$ represents the constant stiffness of the harmonic potential. Here we have set the mass and the charge of the particle to be unit for simplicity. The magnetic field $B(t)$ is an odd controlling parameter which changes its sign under time-reversal operation. The motion of the particle is governed by the Langevin equation
\begin{widetext}
\begin{eqnarray} && \frac{\mathrm{d}x}{\mathrm{d}t}=p_{x} +\frac{B}{2}y, \quad     \frac{\mathrm{d}p_{x}}{\mathrm{d}t} = \frac{B}{2}\left( p_{y} -\frac{B}{2}x   \right)-kx - \gamma \left( p_{x} +\frac{B}{2}y   \right) + \xi_{x}(t), \nonumber \\ && \frac{\mathrm{d}y}{\mathrm{d}t}=p_{y} -\frac{B}{2}x, \quad     \frac{\mathrm{d}p_{y}}{\mathrm{d}t} = - \frac{B}{2}\left( p_{x} +\frac{B}{2}y   \right)-ky - \gamma \left( p_{y} -\frac{B}{2}x   \right) + \xi_{y}(t),  \label{app:chargedLE}\end{eqnarray}
where $\xi_{x}(t)$ and $\xi_{y}(t)$ denote the standard Gaussian white noise along the $x$ and $y$ directions, respectively. The noise satisfies the relation:
\begin{equation}   \langle   \xi_{i}(t)    \rangle =0, \quad  \langle   \xi_{i}(t)  \xi_{j}(t')  \rangle    = 2\gamma T\delta_{ij} \delta(t-t') ,  \quad i,j=x,y .         \label{app:chargednoise}\end{equation}
The ensemble behavior of the particle is described by the Fokker-Planck equation
\begin{eqnarray}\frac{\partial \rho}{\partial t} =  \frac{\partial }{\partial x} (D_{1x} \rho) + \frac{\partial }{\partial y} (D_{1y} \rho) + \frac{\partial }{\partial p_{x}} (D_{2x} \rho) + \frac{\partial }{\partial p_{y}} (D_{2y} \rho) + \frac{\partial^{2} }{\partial p^{2}_{x}} (D_{3x} \rho) + \frac{\partial^{2} }{\partial p^{2}_{y}} (D_{3y} \rho),\label{app:chargedfkeq}\end{eqnarray}
where
\begin{eqnarray} && D_{1x}=-p_{x} -\frac{B}{2}y, \quad  D_{1y}=-p_{y} +\frac{B}{2}x , \quad  D_{2x}= -\frac{B}{2}\left( p_{y} -\frac{B}{2}x   \right)+kx + \gamma \left( p_{x} +\frac{B}{2}y   \right) , \nonumber \\ && D_{2y}= \frac{B}{2}\left( p_{x} +\frac{B}{2}y   \right)+ky + \gamma \left( p_{y} -\frac{B}{2}x   \right) ,\quad D_{3x} =D_{3y} = \gamma T.  \label{app:chargedcoeff}\end{eqnarray}

Through steps similar to those in Appendix~\ref{sec:shorttime}, we can derive the corresponding short-time conditional probability
\begin{eqnarray}\rho_{+}(x',y',p_{x}',p_{y}',t+\mathrm{d} t| x,y,p_{x},p_{y},t)   =&&  \delta (x'-x+ D_{1x}\mathrm{d} t )\frac{1}{\sqrt{4 \pi D_{3x} \mathrm{d} t }} e^{-\frac{(p_{x}'-p_{x}+D_{2x} \mathrm{d} t  )^{2}}{4D_{3x}\mathrm{d} t }} \nonumber \\  && \times \delta (y'-y+ D_{1y}\mathrm{d} t )\frac{1}{\sqrt{4 \pi D_{3y} \mathrm{d} t }} e^{-\frac{(p_{y}'-p_{y}+D_{2y} \mathrm{d} t  )^{2}}{4D_{3y}\mathrm{d} t }}
 . \label{app:chargedstp}\end{eqnarray}
Similarly, the short-time conditional probability of the time-reversed driving process can be derived as
\begin{eqnarray}\rho_{-}(\bar{x}',\bar{y}',\bar{p}_{x}',\bar{p}_{y}',\bar{t}+\mathrm{d} \bar{t}| \bar{x},\bar{y},\bar{p}_{x},\bar{p}_{y},\bar{t})  =&&  \delta (\bar{x}'-\bar{x}+ \bar{D}_{1x}\mathrm{d} \bar{t} )\frac{1}{\sqrt{4 \pi \bar{D}_{3x} \mathrm{d} \bar{t} }} e^{-\frac{(\bar{p}_{x}'-\bar{p}_{x}+\bar{D}_{2x} \mathrm{d} \bar{t}  )^{2}}{4\bar{D}_{3x}\mathrm{d} \bar{t} }} \nonumber \\  && \times \delta (\bar{y}'-\bar{y}+ \bar{D}_{1y}\mathrm{d} \bar{t} )\frac{1}{\sqrt{4 \pi \bar{D}_{3y} \mathrm{d} \bar{t} }} e^{-\frac{(\bar{p}_{y}'-\bar{p}_{y}+\bar{D}_{2y} \mathrm{d} \bar{t}  )^{2}}{4\bar{D}_{3y}\mathrm{d} \bar{t} }}
  \label{app:chargedrstp}\end{eqnarray}
with
\begin{eqnarray} && \bar{D}_{1x}=-\bar{p}_{x} -\frac{\bar{B}}{2}\bar{y}, \quad  \bar{D}_{1y}=-\bar{p}_{y} +\frac{\bar{B}}{2}\bar{x} , \quad  \bar{D}_{2x}= -\frac{\bar{B}}{2}\left( \bar{p}_{y} -\frac{\bar{B}}{2}\bar{x}   \right)+k\bar{x} + \gamma \left( \bar{p}_{x} +\frac{\bar{B}}{2}\bar{y}   \right) , \nonumber \\ && \bar{D}_{2y}= \frac{\bar{B}}{2}\left( \bar{p}_{x} +\frac{\bar{B}}{2}\bar{y}   \right)+k\bar{y} + \gamma \left( \bar{p}_{y} -\frac{\bar{B}}{2}\bar{x}   \right) ,\quad \bar{D}_{3x} =\bar{D}_{3y} = \gamma T.  \label{app:chargedrcoeff}\end{eqnarray}
The mapping relation between the variables in the time-reversed driving process and those in the forward driving process follows relation~(\ref{eq:undertimereversedvariables}).
\end{widetext}

The stationary distribution of this system can be derived from the Fokker-Planck equation~(\ref{app:chargedfkeq}) as
\begin{equation}\rho_{\mathrm{st}}(x,y,p_{x},p_{y},B) = \mathrm{e}^{ \frac{1}{T}\left [ F- H(x,y,p_{x},p_{y},B) \right]},\label{eq:chargedequilibrium}\end{equation}
where
\begin{eqnarray}F(\lambda_{1},\lambda_{2}) && \equiv -T\ln \left[\iiiint \mathrm{e}^{ \frac{1}{T}\left ( F- H \right)} \mathrm{d}x \mathrm{d}y\mathrm{d}p_{x} \mathrm{d}p_{y} \right]\nonumber \\ &&= -T\ln \frac{4\pi^{2}}{\beta^{2}k}.\label{eq:chargedfreeenergy}\end{eqnarray}
The stationary distribution~(\ref{eq:chargedequilibrium}) possesses the form of canonical distribution. Combining the stationary distribution~(\ref{eq:chargedequilibrium}) with the short-time conditional probability~(\ref{app:chargedstp}) and~(\ref{app:chargedrstp}), we can verify that the principle of detailed balance is also satisfied
\begin{widetext}
\begin{eqnarray}&&\rho_{\mathrm{st}}(x,y,p_{x},p_{y},B) \rho_{+}(x',y',p_{x}',p_{y}',t+\mathrm{d} t| x,y,p_{x},p_{y},t) \nonumber \\&&= \rho_{\mathrm{st}}(\bar{x},\bar{y},\bar{p}_{x},\bar{p}_{y},\bar{B}) \rho_{-}(\bar{x}',\bar{y}',\bar{p}_{x}',\bar{p}_{y}',\bar{t}+\mathrm{d} \bar{t}| \bar{x},\bar{y},\bar{p}_{x},\bar{p}_{y},\bar{t}). \label{eq:chdetailedsymmetry}\end{eqnarray}
\end{widetext}

Substituting Eqs.~(\ref{app:chargedstp}) and~(\ref{app:chargedrstp}) into microscopic reversibility~(\ref{eq:microreversibility}), we can derive that
\begin{eqnarray} \text{\dj} q = && - T \ln \frac{\rho_{+}(x',y',p_{x}',p_{y}',t+\mathrm{d} t| x,y,p_{x},p_{y},t)}{\rho_{-}(\bar{x}',\bar{y}',\bar{p}_{x}',\bar{p}_{y}',\bar{t}+\mathrm{d} \bar{t}| \bar{x},\bar{y},\bar{p}_{x},\bar{p}_{y},\bar{t})}  \nonumber \\ = && \left( \frac{\mathrm{d} p_{x}}{\mathrm{d}t} +\frac{\partial H}{\partial x} \right) \circ \mathrm{d}x + \left( \frac{\mathrm{d} p_{y}}{\mathrm{d}t} +\frac{\partial H}{\partial y} \right) \circ \mathrm{d}y ,\label{eq:chargedmrratioheat}\end{eqnarray}
which recovers the definition of heat. This indicates that the condition of microscopic reversibility is compatible with the definition of heat for systems with the magnetic field as the only odd controlling parameter. Starting from relation~(\ref{eq:chargedmrratioheat}), we can further verify that the Jarzynski equality and the second law of thermodynamics are also recovered.

Back to Eq.~(\ref{app:chargedhamiltonian}), the Hamiltonian is an even function of canonical momentum. Thus the odd controlling parameter does not break the time-reversal invariance of the Hamiltonian~(\ref{app:chargedhamiltonian}). This may be the essential difference between the system with the magnetic field as the only odd controlling parameter and the system we considered in the main text.

\end{document}